\documentclass[showpacs,prd,twocolumn,floatfix,nofootinbib,superscriptaddress,preprintnumbers,showkeys]{revtex4-1}
\usepackage{booktabs}
\usepackage{color}
\usepackage{amsfonts}
\usepackage{amsmath}
\usepackage{verbatim}
\usepackage{bigints}
\usepackage{ulem}
\usepackage{siunitx}
\usepackage{multirow}
\usepackage{graphicx}
\usepackage{url}

\newcommand{\ms}{\overline{MS}}

\newcommand{\hm}{h}
\newcommand{\trace}{\operatorname{Tr}}

\newcommand{\pslash}{p\!\!\!/}

\newcommand{\kslash}{k\!\!\!/}

\newcommand{\ei}[1]{\operatorname{Ei}\left(#1\right)}
\newcommand{\erf}[1]{\operatorname{erf}\left(#1\right)}

\newcommand{\gE}{\gamma_{\mathrm{E}}}

\newcommand{\euv}{\epsilon_{\scriptscriptstyle\mathrm{UV}}}
\newcommand{\eir}{\epsilon_{\scriptscriptstyle\mathrm{IR}}}
\newcommand{\oL}{\overline{\Lambda}}
\newcommand{\om}{\overline{m}}

\newcommand{\op}{\overline{p}}
\newcommand{\oz}{\overline{z}}

\begin{document}
\title{Smeared quasidistributions in perturbation theory}

\author{Christopher Monahan}

\email[e-mail: ]{cjm373@uw.edu}

\affiliation{Institute for Nuclear Theory, University of Washington,
Seattle, WA 98195-1550, USA}
\affiliation{New High Energy Theory Center and Department of Physics and 
Astronomy, Rutgers, the State University of New Jersey,
136 Frelinghuysen Road, Piscataway, New Jersey 08854-8019, USA}

\begin{abstract}
Quasi- and pseudodistributions provide a new approach to determining parton distribution functions from first principles' calculations of QCD. Here I calculate the flavor nonsinglet unpolarized quasidistribution at one loop in perturbation theory, using the gradient flow to remove ultraviolet divergences. I demonstrate that, as expected, the gradient flow does not change the infrared structure of the quasidistribution at one loop and use the results to match the smeared matrix elements to those in the $\ms$ scheme. This matching calculation is required to relate numerical results obtained from nonperturbative lattice QCD computations to light-front parton distribution functions extracted from global analyses of experimental data.
\end{abstract}


\preprint{INT-PUB-17-043}

\date{\today}
\nopagebreak
\maketitle

\section{Introduction}

A central challenge for QCD, the gauge theory of the strong nuclear force, is the direct prediction of hadron structure. In particular, parton distribution functions (PDFs) and generalized parton distributions (GPDs) have, until recently, posed intractable difficulties for first principles' calculations. Although a complete, \textit{ab initio} determination of PDFs and GPDs has yet to appear, a promising new method was recently proposed in Ref.~\cite{Ji:2013dva}. In this approach, PDFs and GPDs are extracted from matrix elements of spatially extended operators between nucleon states at finite momentum. These matrix elements are generally referred to as quasi-PDFs or quasidistributions. Related frameworks have also been proposed; in Refs.~\cite{Ma:2014jla,Ma:2014jga,Ma:2017pxb} quasidistributions were treated as ``lattice cross sections'' from which light-front PDFs can be factorized, and Refs.~\cite{Radyushkin:2016hsy,Radyushkin:2017cyf,Orginos:2017kos} introduced and studied the closely related pseudodistributions.

Quasidistributions are matrix elements of time-local operators and therefore can be directly calculated using lattice QCD \cite{Carlson:2017gpk,Briceno:2017cpo}, in which QCD is discretized on a Euclidean hypercubic lattice. Preliminary nonperturbative results for both quasi- and pseudodistributions are encouraging \cite{Lin:2014zya,Alexandrou:2015rja,Chen:2016utp,Orginos:2017kos,Zhang:2017bzy,Alexandrou:2017qpu}. Moreover, a number of theoretical issues \cite{Li:2016amo,Rossi:2017muf,Carlson:2017gpk} have been clarified or solved, including a proof of the multiplicative renormalization of the spatial Wilson-line operator \cite{Ji:2017oey,Ishikawa:2017faj}, a proof of the factorization of light-front PDFs from quasidistributions \cite{Ma:2014jla,Ma:2014jga}, and a proof that the matrix element extracted from Euclidean correlation functions is identical to that defined through a Lehmann-Symanzik-Zimmermann reduction procedure in Minkowski space \cite{Briceno:2017cpo,Xiong:2017jtn,Ji:2017rah}. 

One of the computational challenges that must be addressed is the presence of a power divergence induced in the quasi- and pseudodistributions by the finite lattice regulator. The Wilson-line operator, which defines these distributions, has a divergence that scales exponentially with the length of the Wilson line divided by the lattice spacing, and this divergence must be removed nonperturbatively. Several approaches have been suggested: the authors of Refs.~\cite{Chen:2016fxx} and \cite{Ishikawa:2016znu} proposed removing the power divergence through an exponentiated mass renormalization, and, more recently, regularization invariant momentum subtraction (RI/MOM \cite{Chen:2017mzz} and RI' \cite{Alexandrou:2017huk}) schemes have been suggested as nonperturbative renormalization procedures. In Ref.~\cite{Monahan:2016bvm}, we introduced an alternative approach, the smeared quasidistribution, which circumvents some of the challenges of the power divergence by taking advantage of the properties of the gradient flow \cite{Narayanan:2006rf,Luscher:2011bx,Luscher:2013cpa}.

The gradient flow is a classical evolution, or one-parameter mapping, of the original quark and gluon degrees of freedom in a new parameter, the flow time. The flow time exponentially suppresses UV field fluctuations, which corresponds to smearing out the original degrees of freedom in real space. The critical property of the gradient flow is that, up to a multiplicative fermion wave function renormalization \cite{Luscher:2013cpa}, finite correlation functions of the original theory remain finite at nonzero flow time \cite{Luscher:2011bx}. By fixing the flow time in physical units, one can ensure that matrix elements determined nonperturbatively at finite lattice spacing remain finite in the continuum limit. The gradient flow therefore provides a nonperturbative, gauge-invariant method to render the quasidistribution finite, even in the continuum \cite{Monahan:2016bvm}. The resulting continuum matrix element can then be related directly to the light-front PDF, or to the quasi or pseudodistributions renormalized in, for example, the $\ms$ scheme. In practice, it is simpler to carry out this matching procedure in several steps: first relate the smeared distribution to the distribution without smearing, and then use the known relation between the unsmeared distribution and the light-front PDF \cite{Xiong:2013bka,Ji:2015jwa,Wang:2017qyg,Stewart:2017tvs}. The perturbative calculation of the relevant matrix elements in the $\ms$ scheme appears in Ref.~\cite{Constantinou:2017sej}, and here my focus is the determination of the matrix elements at finite flow time.

I study the matrix element that defines the smeared quasidistribution at one loop in perturbation theory. Computing the matrix element of the smeared Wilson-line operator between external massless quark states at rest, enables all integrals to be evaluated analytically. At one loop, the Wilson-line power divergence manifests as a contribution linear in $\oz = z/r_\tau$, where $z$ is the length of the Wilson line and $r_\tau$ is the gradient flow smearing radius, for $\oz \gg 1$. Subtracting this contribution, the remaining matrix element is finite, and has a well-defined $\oz\to0$ limit. In the small flow-time regime, for which $\oz \gg 1$, the matrix element depends only logarithmically on $\oz$ and satisfies a relation analogous to the usual renormalization-group equation.

I start by briefly reviewing light-front PDFs, smeared quasidistributions and their relation in Sec.~\ref{sec:defns}. I discuss the perturbative calculation of the relevant matrix element in Sec.~\ref{sec:pertth}, provide some numerical results in Sec.~\ref{sec:numres}, and summarize in Sec.~\ref{sec:summary}.

\section{\label{sec:defns}Light-front and quasidistributions}

Throughout this work, I focus on flavor nonsinglet unpolarized quasi and light-front PDFs. The extension to polarized quasi-PDFs is straightforward and much of the discussion applies equally to pseudodistributions. I do not consider the flavor singlet case, which introduces additional mixing with the gluon distribution; this mixing is studied for the case of the quasidistribution at one loop in perturbation theory in Ref.~\cite{Wang:2017qyg}. 

\subsection{Light-front PDFs}

I denote renormalized light-front PDFs by $f(\xi,\mu)$, where the renormalization scale is $\mu$, and define $\xi=k^+/P^+$. Here the light-front coordinates are $(x^+,x^-,\boldsymbol{x}_\mathrm{T})$ such that $x^\pm = (t\pm z)/\sqrt{2}$. In general, one can write the renormalized light-front PDFs in terms of the bare light-front PDFs, $f^{(0)}(\xi)$, as
\begin{equation}
f(\xi,\mu) = \int_\xi^1 \frac{\mathrm{d}\zeta}{\zeta}{\cal 
Z}\left(\frac{\xi}{\zeta},\mu\right)f^{(0)}(\zeta).
\end{equation}
The bare PDF is defined as \cite{Collins:2011zzd}
\begin{align}
 {} &f^{(0)}(\xi) = \int_{-\infty}^\infty 
\frac{\mathrm{d}\omega^-}{4\pi}
e^{-i\xi P^+\omega^-} \nonumber\\
{} & \quad \times \left\langle P \left| 
T\,\overline{\psi}(0,\omega^-,\mathbf{0}_\mathrm{T})
W(\omega^-,0)\gamma^+\frac{\lambda^a}{2}\psi(0) \right| 
P\right\rangle_\mathrm{C},
\end{align}
where $T$ is the 
time-ordering operator, $\psi$ is a quark 
field, $\lambda^a$ is an SU(2)-flavor Pauli matrix, and the subscript C indicates that the vacuum 
expectation value has been subtracted (in other words, only connected 
contributions are included). The operator $W(\omega^-,0)$ is the Wilson line,
\begin{equation}
W(\omega^-,0) = 
{\cal P}\exp\left[-ig_0\int_0^{\omega^-}\mathrm{d}y^-A^+_c(0,y^-,
\mathbf{0}_{\mathrm{T}})T_c\right],
\end{equation}
with ${\cal P}$ the path-ordering operator, $g_0$ the QCD bare coupling, 
and $A^\alpha = A^\alpha_c T_c$ the $SU(3)$ gauge potential with generator 
$T_c$ (summation over color index $c$ is implicit). The target 
state, $|P\rangle$, is a spin-averaged, exact momentum eigenstate with 
relativistic normalization
\begin{equation}
\langle P' | P \rangle = 
(2\pi)^32P^+\delta\left(P^+-P^{\prime\,+}\right)\delta^{(2)}
\left(\mathbf{P}_\mathrm{T} - \mathbf{P}'_\mathrm{T}\right).
\end{equation}

\subsection{Smeared quasidistributions}

One of the central challenges for nonperturbative calculations of quasidistributions is a power divergence associated with the length of the Wilson line in the presence of the lattice regulator. A number of approaches have been proposed to remove this power divergence \cite{Chen:2016fxx,Ishikawa:2016znu,Chen:2017mzz,Alexandrou:2017huk,Orginos:2017kos}, including circumventing the issue completely by using alternative operators \cite{Aglietti:1998ur,Abada:2001if,Detmold:2005gg,Braun:2007wv,Bali:2017gfr}. In Ref.~\cite{Monahan:2016bvm}, we constructed a finite matrix element to extract the continuum quasidistribution from lattice calculations by smearing both the fermion and gauge fields via the gradient flow 
\cite{Narayanan:2006rf,Luscher:2011bx,Luscher:2013cpa}. For a more complete discussion of the gradient flow, I refer the reader to the recent reviews 
\cite{Luscher:2013vga,Ramos:2015dla}. 

At finite flow time, the gradient flow guarantees that the matrix elements constructed from flowed fields is UV finite. Provided one holds the flow time fixed in physical units, the lattice regulator can be removed and one can extract a continuum quasidistribution that is a function of the flow time. This smeared quasidistribution can then be related to the light-front distributions through a perturbative matching relation \cite{Monahan:2016bvm}, analogous to the relation that holds for the ordinary (unsmeared) quasidistribution. Here, I study, at one loop in perturbation theory, the matrix element that defines the smeared quasidistribution. This result is necessary to relate the smeared matrix element to the corresponding matrix element in the $\ms$ scheme.

The continuum smeared matrix element is finite but, because of the presence of the flow-time scale, is expected to contain a power divergence as the flow time is taken to zero \cite{Monahan:2016bvm}. This contribution must be removed nonperturbatively. The gradient flow provides a continuous parameter, in contrast to the lattice regulator, which should provide a better lever through which to study the power-divergent contribution. At one loop in perturbation theory, this divergence manifests as a term linear in the length of the Wilson line, an expectation confirmed by the results presented in Sec.~\ref{sec:pertth}.

I denote the ringed fermion fields \cite{Makino:2014taa,Hieda:2016lly} at flow time $\tau$ by $\overline{\chi}(x;\tau)$ and $\chi(x;\tau)$, and the corresponding Wilson line 
at the same flow time, constructed from the smeared gauge fields $B_\alpha(x;\tau)$, by ${\bf\cal W}(x_1,x_2;\tau)$. I start with the connected matrix element
\begin{align}\label{eq:hdef}
{} & \hm^{(s)}\left(\frac{n^2}{\tau},n\cdot P,
\sqrt{\tau}\Lambda_{\mathrm{QCD}},\sqrt{\tau}M_\mathrm{N}\right) = \nonumber\\
{} & \quad 
\frac{1}{2P}\left\langle 
P \left| \overline{\chi}(n;\tau) 
{\bf\cal W}(n,0;\tau)\gamma_\alpha\frac{\lambda^a}{2}\chi(0;\tau)\right| 
P\right\rangle_\mathrm{C},
\end{align}
which, being dimensionless, depends only on dimensionless combinations of scales. Here, $n$ is a 4-vector that is usually taken to be in the $z$-direction, $n = (\mathbf{0},z,0)$; the dependence of the matrix element on 4-vectors is constrained by Euclidean SO(4) invariance \cite{Radyushkin:2016hsy}. I note that the flow time has units of length squared. The ringed fermion fields require no wave function renormalization and this smeared matrix element is finite provided the flow time, $\tau$, is nonzero and fixed in physical units, 
because correlation functions constructed from smeared fields are finite \cite{Luscher:2011bx,Luscher:2013cpa}. Divergences will appear in the 
limit of vanishing flow time and the matrix element will then require renormalization. The Lorentz index $\alpha$ can take any value from 1 to 4, but the choice $\alpha = 4$ removes some higher-twist contamination \cite{Radyushkin:2016hsy} and avoids mixing, at least for the unpolarized flavor-nonsinglet distribution, at finite lattice spacing \cite{Constantinou:2017sej}. In lattice calculations, it is common to choose the spatial momentum of the nucleon state to be $\mathbf{P} = (0,0,P_z)$, in which case the index $\alpha$ is restricted to be $\alpha = \{3,4\}$.

The smeared quasidistribution \cite{Monahan:2016bvm} is then defined, assuming $n = (\mathbf{0},z,0)$, as
\begin{align}\label{eq:qpdfdef}
q^{\,(s)}{} & \left(\xi,\sqrt{\tau}P_z,\sqrt{\tau}\Lambda_{\mathrm{QCD}}, 
\sqrt{\tau}M_\mathrm{N}\right) 
= 
\int_{-\infty}^\infty \frac{\mathrm{d}z}{2\pi} e^{i\xi z 
P_z} P_z\nonumber \\
{} & \qquad \times h^{(s)}\left(\frac{n^2}{\tau},n\cdot P,\sqrt{\tau}\Lambda_{\mathrm{QCD}}, 
\sqrt{\tau}M_\mathrm{N}\right),
\end{align}
where $\xi$ is a dimensionless parameter that in Euclidean space should be viewed as a (dimensionless) Fourier-conjugate variable.  

\subsection{Relating smeared quasidistributions and PDFs}

The smeared quasidistribution can be directly related to the light-front distribution by \cite{Monahan:2016bvm}
\begin{align}
{} & q^{\,(s)}\left(x,\sqrt{\tau} \Lambda_{\mathrm{QCD}}, \sqrt{\tau}P_z\right) =  \nonumber \\
{} & \; \int_{-1}^{1} 
\frac{d\xi}{\xi}\, \widetilde{Z}\left(\frac{x}{\xi},\sqrt{\tau}\mu, 
\sqrt{\tau}P_z\right) 
f(\xi,\mu) + 
{\cal O}(\sqrt{\tau}\Lambda_{\mathrm{QCD}}),
\end{align}
provided
\begin{equation}\label{eq:scales_tau}
\Lambda_{QCD},M_N\ll P_z \ll \tau^{-1/2}.
\end{equation}
The quasidistribution and the light-front PDF have the same IR structure \cite{Ma:2014jla,Ma:2014jga,Carlson:2017gpk,Briceno:2017cpo}, so that the matching kernel can be determined in perturbation theory. In practice, it is simplest to relate the smeared quasidistribution to the quasidistribution at zero flow time first and then match this (unsmeared) quasidistribution to the light-front PDF \cite{Xiong:2013bka,Ji:2015jwa}.

\begin{figure}
\centering
\includegraphics[width =.25\textwidth,keepaspectratio=true]{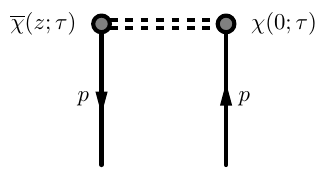}
\caption{Tree-level contribution to the smeared nonsinglet quasidistribution. The gray circles indicate fermion fields at finite flow time $\tau$ and solid lines represent fermion propagators.}
\label{fig:sqpdf_tree}
\end{figure}

\begin{figure}
\centering
\includegraphics[width =0.4\textwidth]{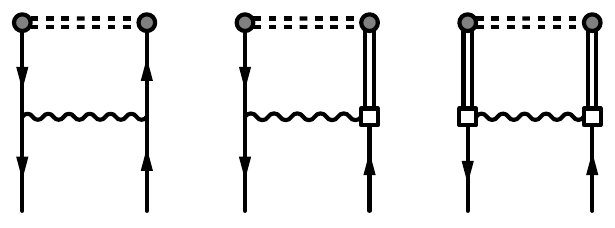}\vspace*{-30pt}\\
(a)\hspace*{0.12\textwidth}(b)\hspace*{0.12\textwidth}(c)\vspace*{10pt}\\
\includegraphics[width =0.42\textwidth]{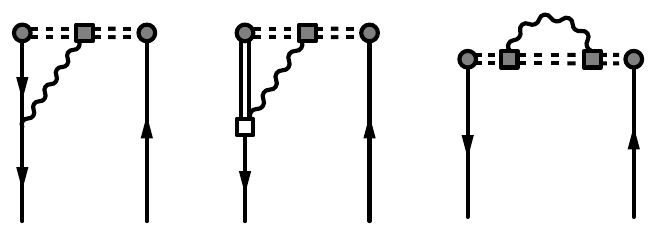}\vspace*{-30pt}\\
\hspace*{-0.02\textwidth}(d)\hspace*{0.12\textwidth}(e)\hspace*{0.12\textwidth}(f)\vspace*{10pt}\\
\includegraphics[width =0.48\textwidth]{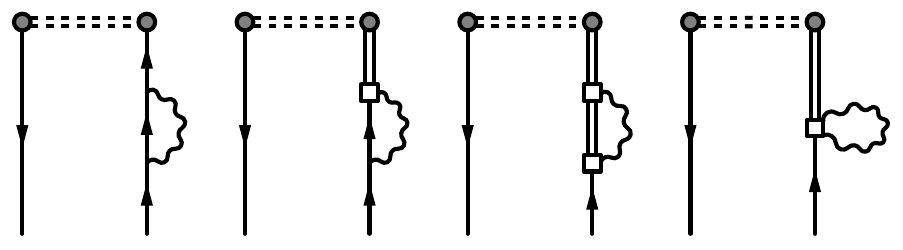}\vspace*{-30pt}\\
\hspace*{-0.02\textwidth}(g)\hspace*{0.1\textwidth}(h)\hspace*{0.1\textwidth}(i)\hspace*{0.1\textwidth}(j)\vspace*{10pt}\\
\caption{ Diagrams representing the one-loop contributions to the smeared nonsinglet quasidistribution. The gray circles indicate fermion fields at finite flow time $\tau$; solid lines represent fermion propagators; double lines are fermion flow kernels; and open squares are flow vertices at arbitrary flow time $\tau_1$. The diagrams labeled (a), (d), (f), and (g) are in one-to-one correspondence with the diagrams that contribute to the unsmeared quasidistribution in the $\ms$ scheme, for which the fermion fields that appear in the extended operator are replaced by fermion fields at vanishing flow time.}
\label{fig:sqpdf1}
\end{figure}

I show the diagrams representing the tree-level and one-loop contributions to the smeared quasidistribution in generalized Feynman gauge (with $\alpha = \lambda = 1$, where $\alpha$ is the usual gauge-fixing parameter and $\lambda$ is a parameter introduced to fix the gauge of the smeared gauge fields \cite{Luscher:2010iy,Luscher:2011bx}) in Figs.~\ref{fig:sqpdf_tree} and \ref{fig:sqpdf1}, respectively. These diagrams correspond to the ``quark-in-quark'' distributions at one loop in perturbation theory. Although the quasidistribution can be more easily evaluated in axial gauge, this gauge cannot be consistently generalized to arbitrary flow time.
In principle, it is possible to evaluate these nine diagrams and determine the matching kernel directly at one loop in perturbation theory. In practice, however, the corresponding integrals cannot be evaluated analytically, and it is simpler to use a procedure similar to that employed in Refs.~\cite{Ishikawa:2016znu,Constantinou:2017sej} and match at a fixed value of the Wilson-line length, $z$. This procedure, which corresponds to matching at the level of the matrix element itself, rather than dealing with the quasidistributions, is preferable largely because the matrix elements do not have any dependence on $\xi$. In the rest of this work, I focus on the matrix element, $\hm^{(s)}$, itself.

\section{\label{sec:pertth}Matrix elements in perturbation theory}

I illustrate the leading-order, or tree-level, contribution to the smeared nonsinglet matrix element in Fig.~\ref{fig:sqpdf_tree}. In both Figs.~\ref{fig:sqpdf_tree} and \ref{fig:sqpdf1}, I use the following conventions: solid lines represent fermion propagators; double lines are fermion flow kernels; open squares are flow vertices at arbitrary flow time $\tau_1$; gray circles indicate fermion fields at fixed flow time $\tau$; gray squares indicate vertices at fixed flow time $\tau$; and the dashed line is the spatially extended Wilson-line operator. 
Note that gauge kernels do not contribute to the smeared nonsinglet quasidistribution at one loop in perturbation theory. The tree-level contribution is given by
\begin{align}
\hm^{(0)} = {} & \frac{1}{2p}\big\langle  \,p,s \,\big| \overline{\chi}(n;\tau)\gamma_\alpha\frac{\lambda^a}{2}\chi(0;\tau) \big|\, p,s \, \big\rangle \nonumber\\
= {} & \frac{1}{2p}e^{ip_zz} e^{-2p^2\tau}\overline{u}^s(p)\frac{\lambda^a}{2}\gamma_\alpha u^s(p),
\end{align}
where I have dropped the arguments of the matrix element for clarity, and I have used 
\begin{equation}
\chi(x;\tau) \big | \,p,s \,\big\rangle = e^{-p^2\tau}\psi(x)\big | \,p,s \,\big\rangle =e^{-p^2\tau}e^{ip\cdot x}u^s(p),
\end{equation}
and an analogous relation for the antiquark.

Beyond tree-level, the operator receives radiative corrections that can be written as
\begin{align}
\hm^{(\mathrm{s})}= {} & {\cal Z}_h \hm^{(0)} \nonumber\\
= {} & Z_h^{(1)}(\oz^2,\mu^2\tau) Z_\chi^{(1)}(\mu^2\tau) \big(Z_\psi^{(1)}\big)^{-1}\hm^{(0)}+{\cal O}(\alpha_s^2). \label{eq:matchrel}
\end{align}
Here, $\alpha_s = g^2/(4\pi)$ is the renormalized coupling constant, which is equal to the bare coupling constant to this order and is most naturally evaluated at the scale $\mu^2 = 1/r_\tau^2$. I show the individual one-loop diagrams that contribute to the correction parameter ${\cal Z}_h$ in Fig.~\ref{fig:sqpdf1}. Diagrams (a) to (f) contribute to the operator renormalization parameter, $Z_h$. The wave function renormalization $Z_\psi$ receives contributions from diagram (g). The extra fermion wave function renormalization $Z_\chi$, which receives contributions from diagrams (h), (i) and (j), is automatically removed via the ringed fermions.

The determination of the operator renormalization parameter, $Z_h$, is the central result of this paper. Both the wave function renormalization parameters appear elsewhere (in particular, $Z_\chi$ is calculated in Refs.~\cite{Luscher:2013cpa,Makino:2014taa}); I calculate these contributions for completeness and as a check of my methods and combine these results with $Z_h$ to obtain the complete one-loop correction ${\cal Z}_h$. I show that, in the small flow-time and local vector current limits, the parameter ${\cal Z}_h$ reduces to the corresponding results in the literature \cite{Makino:2014taa,Hieda:2016lly}, a nontrivial check of my results. The existence of a well-defined local vector-current limit for the extended operator is in contrast to, for example, the extended operator in the $\ms$ scheme in the absence of the flow time \cite{Constantinou:2017sej} and is one attractive feature of the smeared quasidistribution.

I split the following discussion into two parts, starting with an analytic study of the UV structure of the matrix elements. This is sufficient to extract $Z_h^{(1)}$ and is possible because I assume the renormalization scale is much larger than the external momentum and set the momentum to zero \cite{Ishikawa:2016znu}. Diagrams (b), (c), (e), and (i) are proportional to the external momentum and therefore vanish in this case. Moreover, the remaining integrals themselves are considerably simplified and become analytically tractable. I regulate the resulting IR divergences with dimensional regularization and demonstrate that these divergences cancel in the matching procedure between the matrix elements in the $\ms$ scheme and at nonzero flow time. In the second part of my discussion, Sec.~\ref{sec:numres}, I numerically evaluate all nine flow-time-dependent diagrams as a function of the external momentum, quark mass, and flow time to illustrate the behavior of the full matrix elements at one loop in perturbation theory.

\subsection{Vertex-type diagrams}

At one loop in perturbation theory, there are three nonvanishing ``vertex-type'' diagrams associated with the extended operator, which are diagrams (a), (d), and (f) in Fig.~\ref{fig:sqpdf1}, and four ``wave-function-type'' diagrams, which are diagrams (g) to (j) in Fig.~\ref{fig:sqpdf1}. At finite Wilson-line length, $z$, the diagrams exhibit the following UV and collinear behavior: diagram (a) is UV finite and logarithmically IR divergent, while diagrams (d) and (f) would be logarithmically UV divergent without the presence of the flow time, which renders them UV finite and IR finite. Thus, I evaluate diagrams (d) and (f) directly in four dimensions, but diagram (a) requires dimensional regularization to regularize the IR behavior of the integral; I choose $d = (4+2\eir)$.  At zero external momentum, the symmetry and dimension of the bare matrix element ensures that the final result is only a function of $\oz^2=z^2/r_\tau^2$ \cite{Radyushkin:2016hsy}, although individual diagrams may also depend on dimensionless combinations of $\mu^2$, $z^2$ and $\tau$. I do not impose any constraints on the functional dependence of any intermediate results, so establishing that the final result depends only on $\oz^2$ is a useful cross-check of the calculation.

To illustrate the calculation, I will consider diagram (a). The calculation of diagrams (d) and (f) proceeds in a very similar manner. At zero external momentum, the contribution to $Z_h$ from this diagram is given by
\begin{equation}
Z^{\mathrm{(a)}} = C_F (-ig_0\gamma_\rho)\! \int_k \frac{e^{-\tau k^2}}{i\kslash}\gamma_\alpha e^{-ik\cdot n} \frac{e^{-\tau k^2}}{i\kslash}(-ig_0\gamma_\sigma)\frac{\delta_{\rho\sigma}}{k^2}.
\end{equation}
Here $C_F = 4/3$ is the color factor; $g_0$ is the bare coupling; all repeated Lorentz indices are summed over; and I define
\begin{equation}
\int_k = \mu^{4-d}\int \frac{\mathrm{d}^dk}{(2\pi)^d}.
\end{equation}

The Lorentz structure of the numerator is straightforward and, using the Schwinger representation for the denominator, this integral becomes
\begin{align}\label{eq:hma}
Z_h^{\mathrm{(a)}} = {} &  g_0^2C_F \frac{\mu^{4-d}}{(2\pi)^d} \int_0^\infty\mathrm{d}\alpha\,\frac{\alpha}{2} \nonumber \\
& \times \int \mathrm{d}^dk\, N_\alpha(k,d)e^{-(2\tau+\alpha) k^2} e^{-ik\cdot n}.
\end{align}
The numerator $N(k,d)$ depends, at finite flow time and spatial separation, on the choice of Lorentz structure in the original matrix element. Introducing momentum components parallel and perpendicular to the spacelike separation, $k_z$ and $k_\perp=(k_x,k_y,k_t)$, respectively, this numerator is given by
\begin{equation}
N(k,d) \equiv (2-d)(k_z^2-k_\perp^2)
\end{equation}
for the choice $\gamma_\alpha = \gamma_z$ in $\hm^{(s)}$ in Eq.~\eqref{eq:hdef} and
\begin{equation}
N(k,d) \equiv (2-d)\left(\displaystyle\frac{3-d}{d-1}k_\perp^2 - k_z^2\right)
\end{equation}
otherwise. 

The momentum integral in Eq.~\eqref{eq:hma} can be carried out by integrating over $k_z$ and $k_\perp$ individually, leaving a Schwinger parameter integral over $\alpha$ that can be done analytically, giving
\begin{align}
Z_h^{\mathrm{(a)}} = {} & \left(\frac{g_0}{4\pi}\right)^2C_F\bigg[\frac{1}{\eir} -\gE-\log\left(\pi \mu^2z^2\right) +3\nonumber\\
{} - \frac{3}{\oz^4} & \Big(e^{-\oz^2}-1\Big)- \frac{1}{\oz^2} \left(4-e^{-\oz^2}\right)+\ei{-\oz^2}\bigg], \label{eq:hta1} 
\end{align}
and
\begin{align}
Z_h^{\mathrm{(a)}} = {} &\left(\frac{g_0}{4\pi}\right)^2C_F \bigg[\frac{1}{\eir}-\gE-\log\left(\pi \mu^2z^2\right)+1\nonumber\\
&{} -\frac{1}{\oz^4}\Big(1-e^{-\oz^2}(1+\oz^2)\Big)+\ei{-\oz^2} \bigg], \label{eq:hta2}
\end{align}
respectively. Here, $\ei{z}$ is the exponential integral
\begin{equation}
\ei{z} = -\int_{-z}^\infty \frac{e^{-t}}{t}\mathrm{d}t.
\end{equation}
As one would expect, the gradient flow does not affect the IR structure of this diagram (see the Appendix for the equivalent result without the presence of the gradient flow).

I treat the other diagrams, (d) and (f), in a similar fashion and find
\begin{align}
Z_h^\mathrm{(d)} = {} & 2 g_0^2C_F\frac{\pi^2}{(2\pi)^4}\int_0^\infty\mathrm{d}\alpha\,\frac{\alpha}{(\alpha+2\tau)^2}\nonumber \\
{} & \qquad \qquad\qquad\times \left(1-e^{-z^2/(4(\alpha+2\tau))}\right),\\
Z_h^\mathrm{(f)} = {} & g_0^2C_F \frac{\pi^{d/2}}{(2\pi)^4}\int_0^\infty\mathrm{d}\alpha\,\frac{1}{(\alpha+2\tau)^{3/2}} \nonumber \\
\times {} & \int_0^\infty\mathrm{d}\beta\frac{1}{\sqrt{\alpha+\beta+2\tau}}\left(e^{-z^2/(4(\alpha+\beta+2\tau))}-1\right).
\end{align}
The results are
\begin{align}
\!\!Z_h^\mathrm{(d)} = {} & -\left(\frac{g_0}{4\pi}\right)^2C_F \cdot2\bigg[1-\gE-\log\left(\oz^2\right)+\ei{-\oz^2} \nonumber \\
{} & \qquad  + \frac{1}{\oz^2}\left(e^{-\oz^2}-1\right)\bigg], \label{eq:htd}\\
\!\!
Z_h^\mathrm{(f)} = {} & \left(\frac{g_0}{4\pi}\right)^2C_F \cdot2\bigg[\gE+\log\left(\oz^2\right)- \ei{-\oz^2}  \nonumber \\
{} & \quad {-} 2\left(e^{-\oz^2}-1\right) -2\sqrt{\pi\oz^2}\erf{\sqrt{\oz^2}}\bigg],
\label{eq:htf}
\end{align}
where $\erf{z}$ is the error function
\begin{equation}
\erf{z} = \frac{2}{\sqrt{\pi}}\int_0^z e^{-t^2}\mathrm{d}t.
\end{equation}
This term is the one-loop contribution to the expected power divergence and scales linearly with the length of the Wilson line for $\oz \gg 1$.

\subsection{Wave-function-type diagrams}

The ``wave-function-type'' diagrams are diagrams (g), (h), and (j) in Fig.~\ref{fig:sqpdf1} [diagram (i) is proportional to the external momentum]. The contribution from diagram (g) is just the standard quark wave function renormalization, which will cancel exactly in the one-loop matching relation between the smeared and $\ms$ distributions, because it is independent of the flow time. 
In general, this diagram vanishes in dimensional regularization for vanishing external momentum and in the absence of any other scales, but it is helpful to expose the UV and IR divergences by introducing a separation scale, and I discuss this in more detail in the Appendix.

The divergent pieces of the flow-time-dependent diagrams (h) and (j) are removed through the use of ringed fermions, which incorporates the extra fermion renormalization $Z_\chi$ associated with the fermions at finite flow time. Diagrams (h) and (j) include flow vertices that occur at arbitrary flow times $\tau_i$ and must be integrated from zero to $\tau$. The contributions from these diagrams are represented by
\begin{align}
Z_h^\mathrm{(h)} = {} & 2 g_0^2C_F\cdot 2\int_0^\tau \mathrm{d}\tau_1 
\int_k \frac{e^{-2k^2\tau_1}}{k^2} \nonumber \\
= {} & \left(\frac{g_0}{4\pi}\right)^2  C_F \cdot 2\left[\frac{1}{\euv} 
+1+\log\big(8\pi\mu^2\tau\big)\right], \label{eq:hth}\\
Z_h^\mathrm{(j)}{} & (\mu^2\tau) = {-} 2g_0^2C_F  d\int_0^\tau \mathrm{d}\tau_1\int_k\frac{e^{-2k^2\tau_1}}{k^2} \nonumber \\
= {} & -\left(\frac{g_0}{4\pi}\right)^2C_F \cdot 4\left[\frac{1}{\euv} +\frac{1}{2} + \log(8\pi\mu^2\tau)\right]\label{eq:htj},
\end{align}
where I have included a factor of 2 to account for the corresponding antiquark corrections. These contributions, which are independent of the nonlocal operator structure and are canceled by terms in $Z_\chi$, were first calculated for local quark bilinear operators in Ref.~\cite{Endo:2015iea}.

\subsection{One-loop renormalization parameter}

Drawing together these results, I obtain the one-loop expression for the parameter $Z_h$,
\begin{equation}
Z_h^{(1)} =   1 + \frac{\alpha_s}{3\pi}\Big[Z_h^\mathrm{(a)}+Z_h^\mathrm{(d)} +Z_h^\mathrm{(f)}+Z_h^\mathrm{(g)}+Z_h^\mathrm{(h)}+Z_h^\mathrm{(j)}\Big],
\end{equation}
which is
\begin{align}
Z_h^{(1)} = & {} 1 + \frac{\alpha_s}{3\pi}\bigg[-\frac{3}{\euv}+3\gE+3\log\left(\oz^2\right) -3\operatorname{Ei}(-\oz^2)\nonumber\\
{}  -4&\sqrt{\pi\oz^2}\erf{\sqrt{\oz^2}} -3\log(8\pi\mu^2\tau)+a(\oz^2)\bigg],
\end{align}
where 
\begin{equation}
a(\oz^2) =  5 + \frac{3}{\oz^4} \Big(1-e^{-\oz^2}\Big)- \frac{1}{\oz^2} \left(2+e^{-\oz^2}\right)-4e^{-\oz^2},
\end{equation}
for $\gamma_\alpha = \gamma_z$ and
\begin{equation}
a(\oz^2) =  3 - \frac{1}{\oz^4} \Big(1-e^{-\oz^2}\Big)+ \frac{1}{\oz^2} \left(2-e^{-\oz^2}\right)-4e^{-\oz^2}
\end{equation}
otherwise. These expressions are the central results of this work.

\subsubsection{Asymptotic behavior}
These one-loop expressions satisfy a number of cross-checks, based on the structure expected in the asymptotic regimes $\oz \ll1 $ and $\oz \gg 1$, which correspond to the local vector-current and small flow-time limits, respectively. 

\paragraph{Vector-current limit} In principle, in contrast to the dimensionally regularized case (see the Appendix), there are no contact terms in the limit of vanishing quark-field separation $z\to 0$ \cite{Constantinou:2017sej}, provided one has regulated the IR divergences through an appropriate regulator, such as an IR cutoff or external momentum. In this case, one could directly take the limit $\oz\to 0$. Here, however, I have used dimensional regularization for the IR divergences, which now manifest in a logarithmic dependence on $\mu^2z^2$, and this complicates the procedure. It is simplest to examine the $\oz\ll 1$ regime at the level of each integrand, in which case one finds $Z_h^\mathrm{(d)}=Z_h^\mathrm{(f)}=0$ and
\begin{equation}
\lim_{\oz \to 0}Z_h^\mathrm{(a)} =  \left(\frac{g_0}{4\pi}\right)^2C_F \left[\frac{1}{\eir}+\frac{1}{2} - \log\left(8\pi\mu^2\tau\right)\right],
\end{equation}
independent of the choice of the matrix $\gamma_\alpha$ in the original matrix element, as expected. 

Combining this result with the expressions for diagrams (h) and (j), which are independent of the spatial separation $z$ (but not, of course, the flow time $\tau$), I find
\begin{equation}
\lim_{\oz \to 0}Z_h^{(1)} = \left(\frac{g_0}{4\pi}\right)^2C_F \left\{\frac{1}{2}+\frac{1}{\eir}-\frac{3}{\euv} + \log\left(8\pi\mu^2\tau\right)\right\},\label{eq:hoztozero}
\end{equation}
As I show in the next section, when combined with $Z_\psi$ and $Z_\chi$, the parameter $Z_h^{(1)}$ reduces to the local vector-current result of Ref.~\cite{Hieda:2016lly}, an important check of the calculation. 

\paragraph{Small flow-time limit} In the small flow-time limit, one anticipates that the diagrams reduce to the expressions obtained in the $\ms$ scheme. Diagram (a) reduces to
\begin{equation}
Z_h^{(a)}  \stackrel{\oz \gg 1}{\simeq}\left(\frac{g_0}{4\pi}\right)^2C_F \bigg[\frac{1}{\eir} + C^{(\infty)}_\alpha-\gE-\log(\pi\mu^2z^2) \bigg],
\end{equation}
where $C^{(\infty)}_z = 3$ and $C^{(\infty)}_\perp = 1$. These results are in exact agreement with the $\ms$-scheme expressions in Eq.~\ref{eq:ha_dr}.
Taking $\oz \gg 1$, diagrams (d) and (f) become
\begin{align}
Z_h^\mathrm{(d)}\!\! \stackrel{\oz \gg 1}{\simeq} {} & \!\!\!{-}\left(\frac{g_0}{4\pi}\right)^2C_F 2\bigg[1-\gE-\log(\oz^2) \bigg],\\
Z_h^\mathrm{(f)}\!\! \stackrel{\oz \gg 1}{\simeq} {} & \!\!\!\left(\frac{g_0}{4\pi}\right)^2C_F 2\bigg[1+\gE+\log(\oz^2)  +\left(1-2\sqrt{\pi}\oz\right) \bigg].
\end{align}
The contribution to diagram (f) in parentheses is the linear divergence that must be subtracted before comparison can be made to the calculation carried out with dimensional regularization. Removing this contribution, the sum of these two diagrams is
\begin{equation}
\left(Z_h^\mathrm{(d)} +Z_h^\mathrm{(f)}\right)\stackrel{\oz \gg 1}{=} 4\left(\frac{g_0}{4\pi}\right)^2C_F,\label{eq:hoztoinfty}
\end{equation}
in agreement with the sum of Eqs.~\eqref{eq:hd_dr} and \eqref{eq:hf_dr}. In short, the small flow-time limit of $Z_h$ exactly matches the corresponding result directly calculated in the $\ms$ scheme, as one would expect.

\subsection{One-loop smeared matrix element}

We can now write down the complete smeared matrix element, $\hm^{(\mathrm{s})}$ at one loop in perturbation theory, by combining the wave function renormalization $Z_\psi$ and $Z_\chi$ \cite{Luscher:2013cpa,Makino:2014taa} with the results for $Z_h$ from the previous section. Recalling Eq.~\eqref{eq:matchrel}, the one-loop matrix element is
$\hm^{(\mathrm{s})} = {\cal Z}_hh^{(0)}$, where
\begin{align}
{\cal Z}_h^{(1)} = & {} 1 + \frac{\alpha_s}{3\pi}\bigg[3\log\left(\oz^2\right) +6\gE-3\log(4\pi)-\log(432)\nonumber \\
{} &  -3\operatorname{Ei}(-\oz^2) -4\sqrt{\pi\oz^2}\erf{\sqrt{\oz^2}}+a(\oz^2)\bigg]
\end{align}
and here
\begin{align}
c(\oz^2) = {} & 3\left[3+\frac{1}{\oz^2}\left(e^{-\oz^2}-2\right)+\frac{1}{\oz^4}\left(1-e^{-\oz^2}\right)\right] ,  \label{eq:cmuz}\\
c(\oz^2) = {} & 7 +\frac{1}{\oz^2}\left(3e^{-\oz^2}-2\right) +\frac{1}{\oz^4}\left(e^{-\oz^2}- 1\right),\label{eq:cmuperp}
\end{align}
for $\gamma_\alpha = \gamma_z$ and $\gamma_\alpha = \gamma_\perp$, respectively. I note that the IR divergences in the vertex correction of diagram (a) and the wave function renormalization, diagram (g), have canceled. This matches the IR behavior in the $\ms$ scheme, Eq.~\eqref{eq:zhmsbar}, and demonstrates that the gradient flow does not modify the IR behavior of the quasidistribution. Moreover, the total one-loop result is UV finite (that is, there are no poles in $\euv$), as is guaranteed for nonzero flow time, provided one uses ringed fermions. The logarithmic terms combine in such a way that all dependence on $\mu$ has been eliminated, leaving only $\log(\oz)$. There is a linear divergence generated by the Wilson-line self-energy, diagram (f), that scales as $z/\sqrt{\tau}$, as one would expect on dimensional grounds. At higher orders in perturbation theory, one anticipates that this divergence exponentiates in the form $e^{-cz/\sqrt{\tau}}$.

\begin{figure}
\centering
\caption{Renormalization parameter ${\cal Z}_h^{\mathrm{sub}}$ for $\gamma_\alpha=\gamma_z$ and $\gamma_\alpha = \gamma_\perp$ as a function of $\oz$, where the superscript ``$\mathrm{sub}$'' indicates that the linear divergence has been subtracted. The dot-dashed gray line is ${\cal Z}_h$ including the linear divergence in $\oz$, denoted by ${\cal Z}_h^{\mathrm{div}}$ in the legend. I plot the asymptotic behavior for $\oz \ll 1$ and $\oz \gg 1$, from Eqs.~\eqref{eq:veclimit} and \eqref{eq:zsmalltau} as orange dashed lines.}
\label{fig:Zozplot}
\includegraphics[width=0.47\textwidth,keepaspectratio]{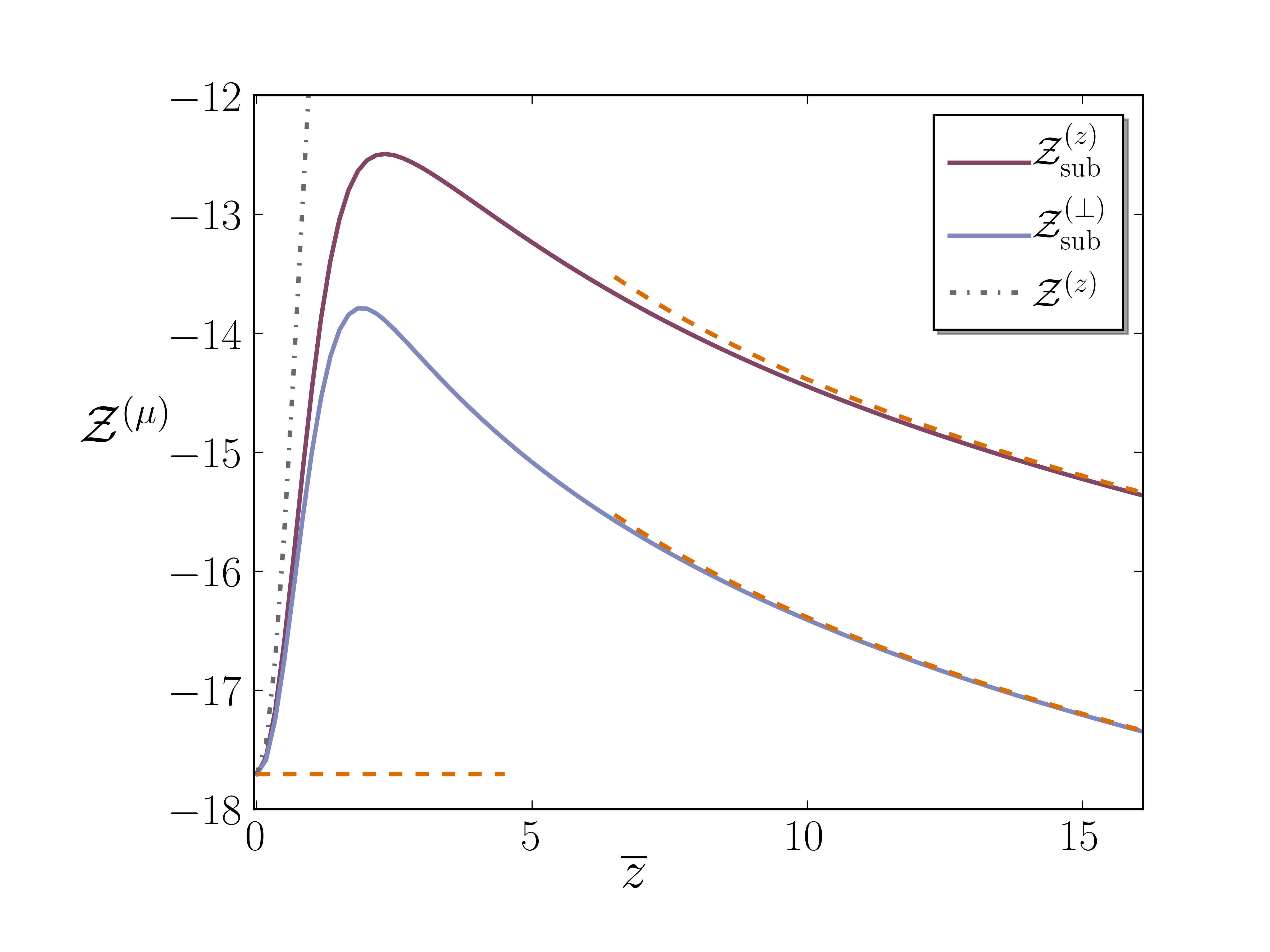}
\end{figure}

I plot both ${\cal Z}_{\mathrm{sub}}^{(z)}(\oz^2)$ and ${\cal Z}_{\mathrm{sub}}^{(\perp)}(\oz^2) $ as functions of $\oz$ in Fig.~\ref{fig:Zozplot}, represented by the purple and blue curves, respectively. I have removed the linear divergence, which would otherwise dominate for values of $\oz\simeq 2$, as illustrated by the dot-dashed gray line, which shows ${\cal Z}^{(z)}(\oz^2)$ including the linear divergence, denoted by ${\cal Z}^{(z)}(\oz^2)$ in the legend. I also plot the asymptotic behavior in the $\oz \ll 1$ and $\oz \gg 1$ regimes as orange dashed lines.

\paragraph{Vector-current limit} In contrast to the matrix element in dimensional regularization, the limit of vanishing spatial separation of the quark fields, $\oz\to 0$, is well defined and is given by
\begin{equation}\label{eq:veclimit}
\lim_{\oz\to0}{\cal Z}_h  = \frac{1}{2}-\log(432),
\end{equation}
independent of the choice one makes for $\gamma_\alpha$ and in exact agreement with the vector-current result of \cite{Hieda:2016lly}. This result exhibits a feature characteristic of composite operators constructed from ringed fermions: perturbative corrections are finite, as they must be for a conserved vector current, but not necessarily zero (compare, for example, with the nonsinglet axial-vector current defined in Ref.~\cite{Endo:2015iea}).

\paragraph{Small flow-time limit} Conversely, the small flow-time expansion gives
\begin{equation}\label{eq:zsmalltau}
{\cal Z}_h^{\mathrm{(sub)}} \stackrel{\oz\gg1}{\simeq} b - \gE - \log(432) - \log(\oz^2),
\end{equation}
with $b = 9$ and $b=7$, for $\gamma_\alpha = \gamma_z$ and $\gamma_\alpha = \gamma_\perp$, respectively. In this regime, the parameter ${\cal Z}_h^{\mathrm{sub}}(\oz^2)$ depends only logarithmically on $\oz$, and therefore satisfies a relation akin to the usual renormalization-group equation \cite{Monahan:2015lha}:
\begin{equation}\label{eq:zsubRG}
\left[\frac{\mathrm{d}}{\mathrm{d}\log\oz^2} +\gamma_{\oz}\right]{\cal Z}_h^{\mathrm{sub}}(\oz^2) = 0.
\end{equation}
Here, $\gamma_{\oz} = \gamma_{\oz}^{(1)}\alpha_s+{\cal O}(\alpha_s^2)$, with $\gamma_{\oz}^{(1)}=1$, is analogous to the usual anomalous dimension. Beyond one loop in perturbation theory, a complete renormalization-group analysis would necessarily take into account the implicit scale dependence of the coupling constant $\alpha_s(\mu)$. A renormalization-group trajectory is then a path in the two-dimensional space $(\oz,\mu)$, which can be simplified by tying the two scales together by choosing, for example, $\mu = 1/r_\tau$ \cite{Monahan:2015lha}. In principle, this could provide an opportunity to construct a nonperturbative step-scaling method for the matrix element $\hm^{(\mathrm{s})}(\oz^2)$, which would connect lattice calculations at hadronic scales to a high scale at which the matching to the $\ms$ scheme can be carried out without significant perturbative truncation uncertainties. Equation \eqref{eq:zsubRG} only holds in the small flow-time regime, however, which, from Fig.~\ref{fig:Zozplot}, applies for $\oz \simeq 10$, at least at one loop in perturbation theory. Imposing this constraint throughout the entire step-scaling procedure may not be feasible for current lattices, but systematic uncertainties associated with deviations from the small flow-time expectation can be studied nonperturbatively.

\subsection{Matching in coordinate space}

With the complete one-loop expression for the smeared quasidistribution in hand, I can match the results to the renormalized matrix element in the $\ms$ scheme, $\hm^{\ms}(\mu^2z^2)$ (see the Appendix). Writing
\begin{equation}
\hm^{\ms}(\mu^2z^2) = {\cal C}_h^{\mathrm{sub}}(\mu^2z^2,\oz^2)\hm^{(\mathrm{s})}(\oz^2),
\end{equation}
the matching factor is given by
\begin{align}
\!\!{\cal C}_h^{\mathrm{sub}}(\mu^2z^2,\oz^2){} &  = 1 +\frac{\alpha_s}{3\pi}\Big[a(\oz^2)-3\overline{c} -\log(432)\nonumber\\ 
 - 3\ei{-\oz^2} {} & -3\log\left(8\pi\mu^2t\right) -4\sqrt{\pi\oz^2}\erf{\sqrt{\oz^2}}\Big].
\end{align}
Here $c(\oz^2)$ is given in Eqs.~\eqref{eq:cmuz} and \eqref{eq:cmuperp}, $\overline{c} =5/3$, and $\overline{c} =7/3$. In general, the linear divergence will be subtracted nonperturbatively and I have removed this contribution from the matching relation, indicated by the subscript ``$\mathrm{sub}$.''
At zero external momentum, this matching parameter is real; at finite external momentum, discussed in the next section, the matching parameter becomes imaginary  \cite{Constantinou:2017sej}.

\section{\label{sec:numres}Numerical results}

The central aim of this work is to study the matrix element $h_\alpha^{(s)}$ at one loop in perturbation theory, and thereby relate the smeared matrix element to the unsmeared matrix element in the $\ms$ scheme. 

In general, relating two renormalized operators can be done with (massless) external quark states with vanishing spatial momentum, provided that momentum does not provide the renormalization scale (such as in the RI/MOM scheme). For completeness, however, I would like to explore the momentum structure of the smeared matrix element in perturbation theory. In this section, I therefore briefly study the momentum-dependent finite pieces of $h_\alpha^{(s)}$, which were not necessary for the matching relation of the previous section.

\paragraph{Analytic approaches} Before I present some sample results, I briefly comment on the calculation at finite external momentum. Introducing a nonzero external momentum considerably complicates the calculation. There are four new diagrams, diagrams (b), (c), (e), and (i), but this increase in the number of diagrams is not too onerous at one loop. More challenging are the momentum integrals themselves. The key difficulty is the exponential damping factor provided by the gradient flow, which prevents the easy application of the residue theorem to the integrals. Moreover, these damping factors also complicate the use of Feynman parameters, because the momentum combinations that appear in the denominator of the integrand are generally not those that appear as arguments of the exponentials, so that one cannot reduce the integrands to radially symmetric integrands. The exponential factors naturally lead one to introduce Schwinger representations of the propagators in the diagram, but these generically lead to integrals over positive Schwinger parameters that are of the form
\begin{equation}
\sum_i\int_0^\infty \prod_j\mathrm{d}\alpha^{(j)}\,f^{(i)}\left(\alpha^{(j)}\right)\,e^{-A^{(i)}\alpha^{(j)}+B^{(i)}/\alpha^{(j)}}.
\end{equation}
Here, the $f^{(i)}$ are polynomials, which may include negative and fractional powers, of the (multiple) Schwinger parameters $\alpha^{(j)}$ (and the physical scales in the problem), and $A^{(i)}$ and $B^{(i)}$ are independent of $\alpha^{(j)}$. These integrals can be carried out numerically but are generally very cumbersome expressions. These integrals seem to be ubiquitous in this approach and also arise from integration over intermediate flow times.

\paragraph{Numeric approach} I determine each momentum integral using the adaptive Monte Carlo \verb+VEGAS+ algorithm \cite{Lepage:1977sw}. I reduce the two-dimensional integral over $\mathbf{k}_\perp$ to a single integral over the positive radial direction in the $(k_x,k_y)$ plane, because the diagrams depend only on $\mathbf{k}_\perp^2$ and not the individual components $k_x$ and $k_y$. I evaluate the Lorentz structure of each integral by hand, and confirm the results using \verb+FeynCalc+ \cite{Mertig:1990an,Shtabovenko:2016sxi}. These manipulations are straightforward at one loop. For simplicity, I take the gamma matrix insertion to be $\gamma_\alpha= \gamma_4$ and express all results in units of $\alpha_s/(3\pi)$. Thus, for the case of diagram (a), I evaluate
\begin{align}
h_4^{(a)} = {} & \frac{1}{2p}g_0^2C_F 2\int_k \frac{  e^{-2\tau k^2} e^{-ik_zz}}{(k^2+m^2)^2(p-k)^2} \nonumber \\
\times \trace {} &\left\{ \left(\frac{-i\pslash+m}{2}\right)\left(4im k_4-2\kslash k_4+ (k^2+m^2)\gamma_4 \right)\right\} \nonumber\\
= {} & \frac{\alpha_s}{3\pi} {\cal I}_4^{(a)},
\end{align}
where the numerical integral is
\begin{align}
{\cal I}_\alpha^{(a)} = {} & \frac{1}{2\op}\frac{8i}{\pi } \int_0^{\oL} \mathrm{d}\ell\,\ell_\perp \int_{-\oL}^{\oL} \mathrm{d}\ell_4\mathrm{d}\ell_z \,e^{-\ell^2} e^{-i\ell_z\oz} \nonumber \\
{} & \qquad \times \frac{2(\op\cdot \ell  +2\om^2)\ell_\alpha-(\ell^2+\om^2)\op_\alpha}{(\ell^2+\om^2)^2(\op-\ell)^2}.
\end{align}
The dimensionless integration variable is $\ell = kr_\tau = k\sqrt{8\tau}$, and the bar indicates the dimensionless combinations $\om = mr_\tau$, $\op = pr_\tau$, and $\oz = z/r_\tau$.

\begin{figure}
\centering
\caption{The real (upper panel) and imaginary (lower panel) parts of the smeared matrix element at one loop, $h_4^{(s)}$, as a function of the spatial Wilson-line length, $z/r_\tau$, for three different values of the external quark momentum, $P_zr_\tau = \{1,2\}$. I take $m_qr_\tau=0.001$ and $P_z/\Lambda = 2$. For these results, I use 20 iterations of $1.5\times 10^8$ measurements following 10 thermalization iterations of $10^6$ measurements each. The smooth curves are interpolations, not fits, to ease comparison between data at different momenta, and the statistical uncertainties are too small to be visible at this scale.}
\label{fig:diagramsum}
\includegraphics[width=0.4\textwidth,keepaspectratio]{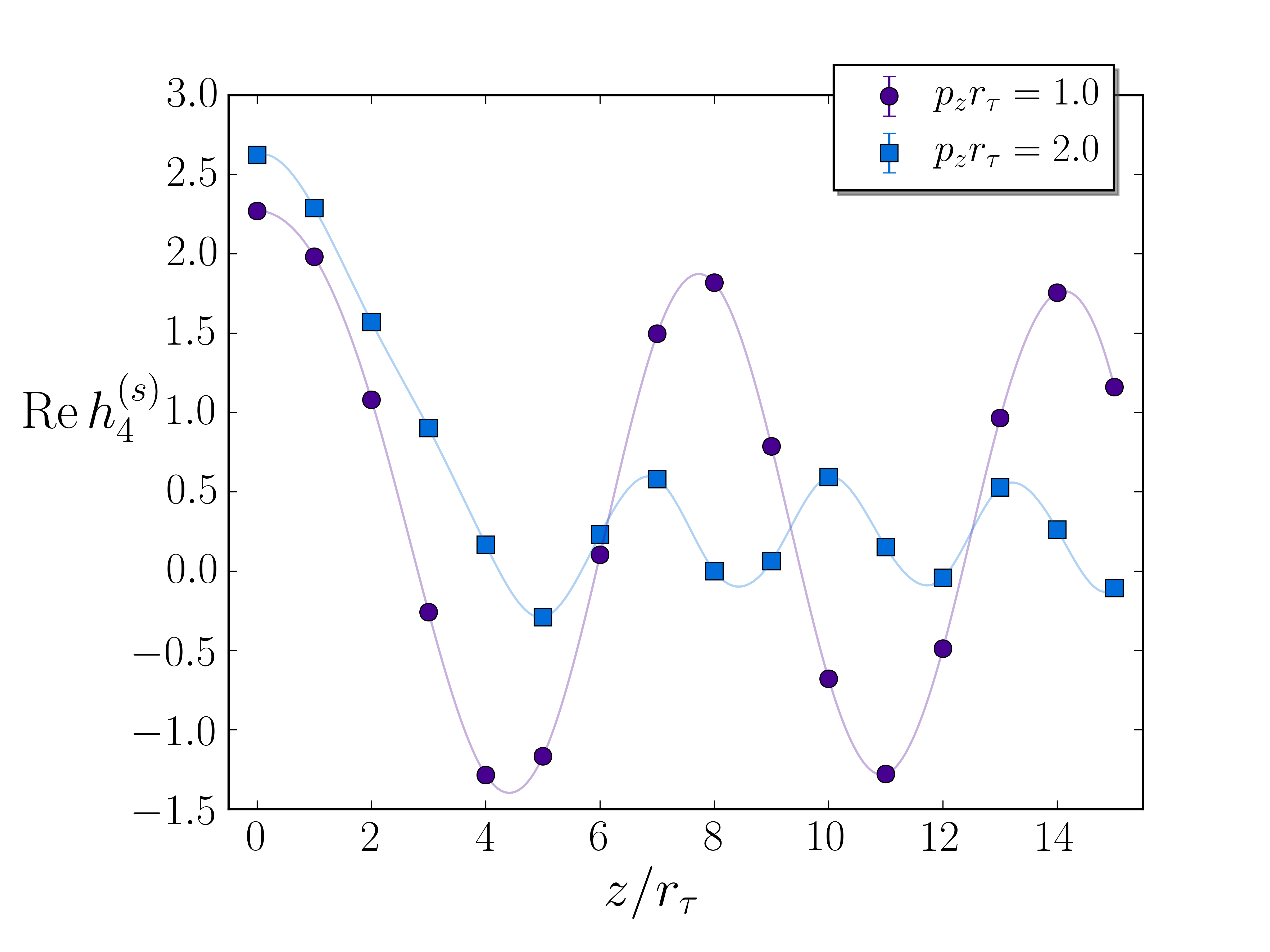}\\
\includegraphics[width=0.4\textwidth,keepaspectratio]{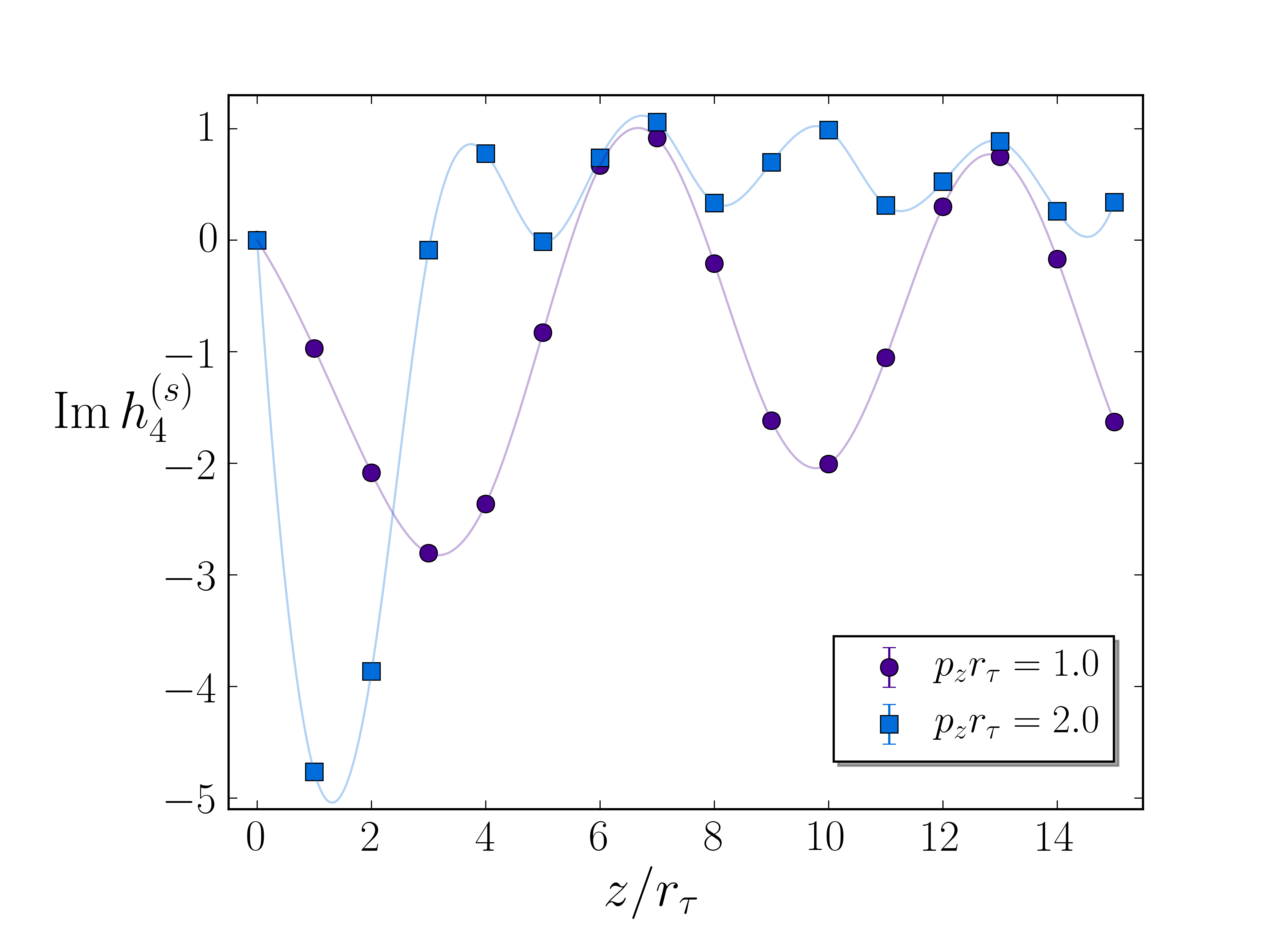}
\end{figure}

In Fig.~\ref{fig:diagramsum}, I provide numerical results over an appropriate range of parameters, based on typical scales in units of the smearing radius. Current lattice calculations use typical lattice spacings of approximately $a\sim \SI{0.1}{fm}\simeq \SI{0.5}{GeV^{-1}}$ and momenta around $p_z \sim 2-\SI{3}{GeV}$. Typical flow times are of the order $r_\tau/a \sim 1$ \cite{Berkowitz:2017opd}---one advantage of the gradient flow is that, in practice, it seems relatively little smearing is required---and so it is reasonable to take $z/r_\tau \sim 0-15$ and $p_zr_\tau\sim 1-2$ as illustrative values.

I plot the real and imaginary parts of the smeared matrix element, as a function of $z/r_\tau$ and for two different values of the quark momentum, in the upper and lower panels of Figs.~\ref{fig:diagramsum}, respectively. The lines are smooth interpolations to the data, not fits, to ease comparison between data points at the two different momenta. The statistical uncertainties from the numerical integration are approximately $\sigma_{\mathrm{VEGAS}}\sim 10^{-4}$, which are too small to be visible at this scale.

\section{\label{sec:summary}Summary}

Quasi and pseudodistributions have opened a new window into nucleon structure and provided an opportunity to extract, for the first time, parton distribution functions directly from lattice QCD. Preliminary results of nonperturbative computations have been encouraging, but there are still challenges to be navigated before the lattice community can claim a robust determination of light-front PDFs. One particular difficulty for lattice QCD is the presence of a power divergence associated with the length of the Wilson line, which must be removed nonperturbatively before one can take a continuum limit. In Ref.~\cite{Monahan:2016bvm}, we introduced the smeared quasidistribution as a tool to circumvent the issue of the power divergence associated with the Wilson line. By taking advantage of the renormalization properties of the gradient flow, a classical evolution of the original degrees of freedom in a new parameter, the flow time, the smeared quasidistribution remains finite in the continuum limit.

Here, I study the smeared quasidistribution at one loop in perturbation theory. Although the gradient flow complicates the perturbative calculation, by introducing new diagrams not present at vanishing flow time and complicating the corresponding integrals, the advantage is significant: provided one fixes the flow time in physical units, the smeared quasidistribution is finite in the continuum limit. The resulting continuum smeared quasidistributions can then be related, directly or through the quasidistribution at zero flow time, to the light-front PDFs in the $\ms$ scheme. The IR behavior of the quasidistribution is unaffected by the flow time, which serves only to regulate the UV behavior, so the matching procedure can be carried out perturbatively, by evaluating the Wilson-line operator between external, gauge-fixed quark states.

I undertake the matching at fixed Wilson-line length, in which case the external quark momentum can be set to zero. This reduces the number of diagrams that contribute at a given order in perturbation theory and simplifies the corresponding one-loop integrals, which can then be carried out analytically. I use dimensional regularization to regularize the logarithmic IR divergences and demonstrate that the resulting poles in the spacetime dimension cancel in the matching relation between the smeared and unsmeared quasidistributions. I show that the limit of vanishing Wilson-line length is well defined and reduces to the known vector current result at finite flow time. In the small flow-time limit, the matrix element depends only logarithmically on the ratio of the Wilson-line length and the gradient flow smearing radius. The smeared matrix element therefore satisfies an equation similar to the typical renormalization-group equation, which provides the potential for defining a nonperturbative step-scaling procedure. Finally, I study the smeared matrix element at finite external momentum by evaluating the corresponding one-loop diagrams numerically. Extending this study to two loops would likely require an automated approach, because the extra vertices induced by the gradient flow cause the number of diagrams to increase rapidly with each order in perturbation theory. A nonperturbative study of the systematic uncertainties associated with the smeared quasi- and pseudodistributions is underway.

\section*{Acknowledgments}
I thank Carl Carlson, Michael Freid, Tomomi Ishikawa, and Christopher Coleman-Smith
for enlightening discussions during 
the course of this work. I am particularly grateful to Kostas Orginos for
numerous insightful conversations and for reading a draft of this manuscript.
I am supported in part
by the U.S.~Department of Energy through Grant No.~DE-FG02-00ER41132.

\appendix

\section{Matrix elements in dimensional regularization}

In this Appendix, I determine the matrix element at vanishing flow time using dimensional regularization to regularize both IR and UV singularities. These results were first calculated in Refs.~\cite{Dorn:1986dt,Chetyrkin:2003vi,Constantinou:2017sej}. I repeat the calculations here because: the intermediate results demonstrate the usefulness of the Schwinger representation of the diagrams; provide a useful cross-check of my new results at nonzero flow time, and demonstrate that both the smeared and unsmeared quasidistributions have the same IR behavior. There are four diagrams that contribute, corresponding to diagrams (a), (d), (f), and (g) in Fig.~\ref{fig:sqpdf1} with the flow time set to zero. The resulting integrals can be reduced to
\begin{align}
Z_h^\mathrm{(a)} = {} & g_0^2C_F \frac{\pi^{d/2}}{(2\pi)^d}\frac{(d-2)^2}{d}\mu^{4-d}\int_0^\infty \frac{\mathrm{d}\alpha}{\alpha^{d/2-1}}  e^{-z^2/(4\alpha)}, \label{eq:ha_int_dr}\\
Z_h^\mathrm{(d)} = {} & 2g_0^2C_F \frac{\pi^{d/2}}{(2\pi)^d}\mu^{4-d} \int_0^\infty \frac{\mathrm{d}\alpha}{\alpha^{d/2-1}}\left(1-e^{-z^2/(4\alpha)}\right), \label{eq:hd_int_dr}\\
Z_h^\mathrm{(f)} = {} & g_0^2C_F \frac{\pi^{d/2}}{(2\pi)^d}\mu^{4-d}\int_0^\infty\frac{\mathrm{d}\alpha}{\alpha^{d/2-1}} \int_0^\infty\frac{\mathrm{d}\beta}{\sqrt{\alpha+\beta}}\nonumber \\
{} & \qquad \times \left(e^{-z^2/(4(\alpha+\beta))}-1\right), \\
Z_h^\mathrm{(g)}= {} & g_0^2C_F \frac{\pi^{d/2}}{(2\pi)^d}\frac{2(2-d)}{\Gamma(d/2)}\mu^{4-d}\int_0^\infty\frac{\mathrm{d}\alpha}{\alpha^{d/2-1}}. \label{eq:hg_int_dr}
\end{align}

I carry out these integrals analytically [with the Schwinger parametrization, this is a straightforward modification to the analogous smeared integrals, with the replacement $(\alpha+2\tau) \to \alpha$]. I regulate the UV divergences in diagrams (d) and (f) by setting the spacetime dimension to $d = (4-2\euv)$ and the IR divergence in diagram (a) by setting $d = (4+2\eir)$. By the rules of dimensional regularization, diagram (g) should vanish, but I wish to expose the UV and IR divergences. I therefore split the Schwinger parameter integral into two regions
\begin{equation}
\int_0^\infty\mathrm{d}\alpha \,\alpha^{1-d/2} = \int_0^\lambda \mathrm{d}\alpha \,\alpha^{1-d/2}+\int_\lambda^\infty\mathrm{d}\alpha \,\alpha^{1-d/2},
\end{equation}
and set $d = (4-2\euv)$ in the UV region, $[0,\lambda]$, and $d = (4+2\eir)$ in the IR region. The result is independent of the choice of $\lambda$.

Then, expanding the results around $\euv = \eir = 0$, I have
\begin{align}
Z_h^\mathrm{(a)} = {} & \left(\frac{g_0}{4\pi}\right)^2C_F \left[\frac{1}{\eir} -\gE+\widetilde{C}_\alpha - \log(\pi\mu^2z^2)\right],\label{eq:ha_dr}\\
Z_h^\mathrm{(d)} = {} & \left(\frac{g_0}{4\pi}\right)^2C_F 2\left[\frac{1}{\euv} +\gE + \log(\pi\mu^2z^2)\right], \label{eq:hd_dr}\\
Z_h^\mathrm{(f)} = {} & \left(\frac{g_0}{4\pi}\right)^2C_F2\left[\frac{1}{\euv}+2 +\gE + \log(\pi\mu^2z^2)\right]. \label{eq:hf_dr}\\
Z_h^\mathrm{(g)} = {} & -\left(\frac{g_0}{4\pi}\right)^2C_F\left[\frac{1}{\euv}+\frac{1}{\eir}\right]. \label{eq:hg_dr}
\end{align}
Here $\widetilde{C}_z = 3$ and $\widetilde{C}_\perp =1$. As one would anticipate, there is no term proportional to $z$, because power-divergent contributions are explicitly removed by dimensional regularization. Diagram (a) has a pole in $1/\eir$, which is not removed by any renormalization procedure, but is canceled by the IR pole in the wave function renormalization, diagram (g). From these expressions, one sees that the $\ms$-scheme renormalization parameter is
\begin{equation}\label{eq:zhmsbar}
Z_h^{\ms} = 1+\frac{\alpha_s}{3\pi}\frac{3}{\euv},
\end{equation}
in agreement with \cite{Dorn:1986dt,Chetyrkin:2003vi,Constantinou:2017sej}.

\section{Numerical integration}

\begin{figure}
\centering
\caption{Numerical convergence for the evaluation of the real part of diagram (a), for different values of the ratio $p_z/\Lambda$ and for four different values of the (dimensionless) external momentum $P_zr_\tau$. The horizontal bands indicate the corresponding results at the smallest value of $P_z/\Lambda$. I take $m_qr_\tau=0.005$ and $z/r_\tau=10$. For these results, I use 20 iterations of $5\times 10^6$ measurements following 10 thermalization iterations of $10^4$ measurements each. The statistical uncertainties are generally too small to be visible at this scale.}
\label{fig:acutoff}
\includegraphics[width=0.47\textwidth,keepaspectratio]{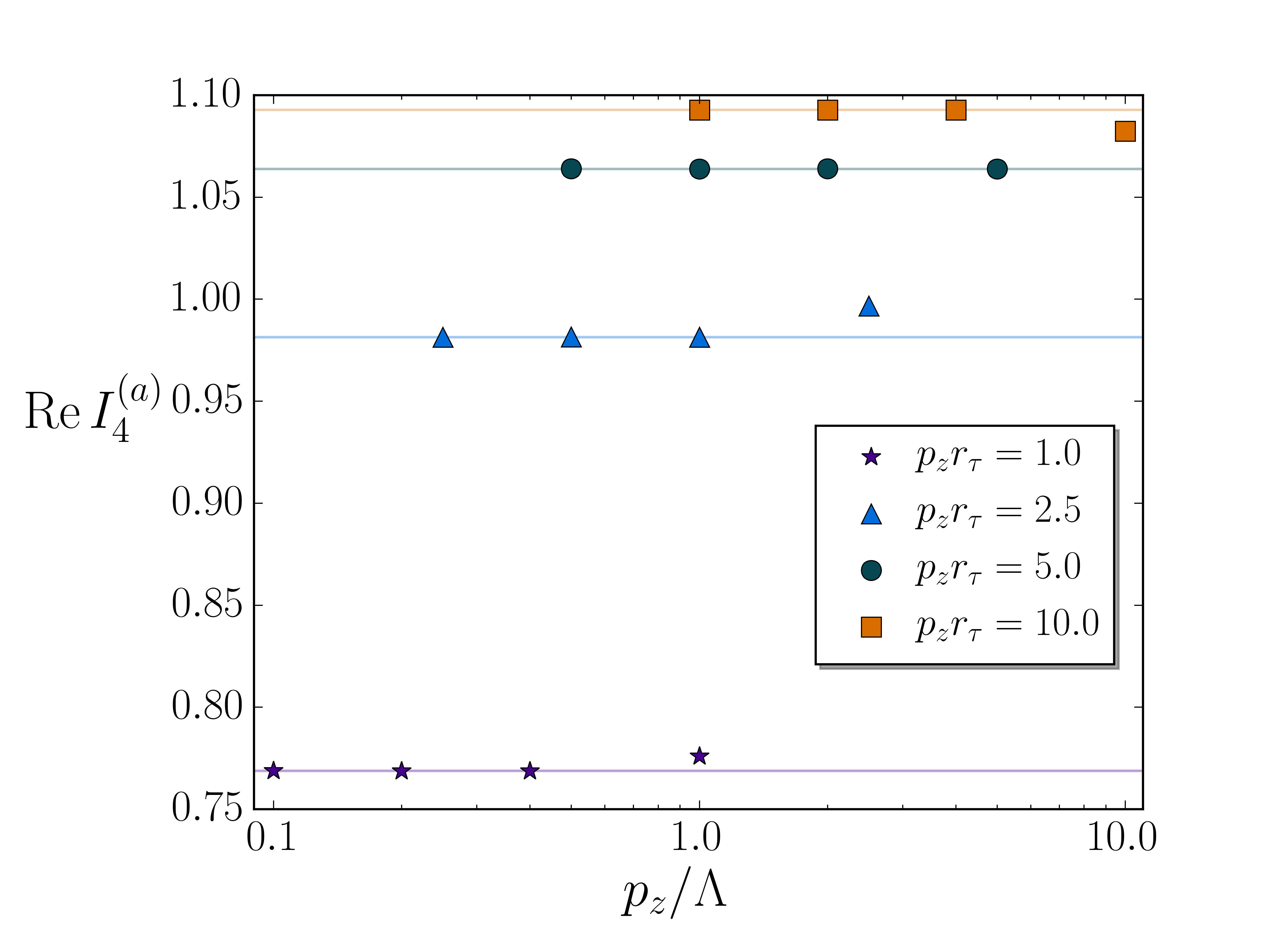}
\end{figure}

In this Appendix, I highlight some details of the numerical integration of Sec.~\ref{sec:numres}, including some tests of the numerical convergence of the results. Once the wave function renormalization contributions are properly incorporated, all diagrams in Fig.~\ref{fig:sqpdf1} are UV and IR finite, so I integrate up to an arbitrary cutoff, $\oL$, chosen to ensure convergence at a given level of numerical precision. I illustrate this convergence in Fig.~\ref{fig:acutoff} for the real part of diagram (a) and observe that, at the current level of precision, convergence occurs for $p_z/\Lambda \simeq 0.5$ for $P_zr_\tau = 1$, and this ratio increases to $p_z/\Lambda = 4$ for $p_zr_\tau = 10.0$. The value of the cutoff at which convergence occurs is diagram dependent, but generally between $p_z/\Lambda \simeq 0.2--0.5$, which ensures that cutoff effects can be neglected, without unnecessarily increasing the statistical uncertainty.

\begin{figure}
\centering
\caption{Numerical convergence for the evaluation of the real (upper plot) and imaginary (lower plot) part of diagram (d). The horizontal bands indicate the corresponding result at $\lambda_{\mathrm{IR}} = 10^{-7}$. I take $m_qr_\tau=0.001$, $z/r_\tau=1$, and $P_z/\Lambda = 2$. For these results, I use 20 iterations of $5\times 10^7$ measurements following 10 thermalization iterations of $10^6$ measurements each, taking approximately 70 sec on three CPU cores.}
\label{fig:dcutoff}
\includegraphics[width=0.4\textwidth,keepaspectratio]{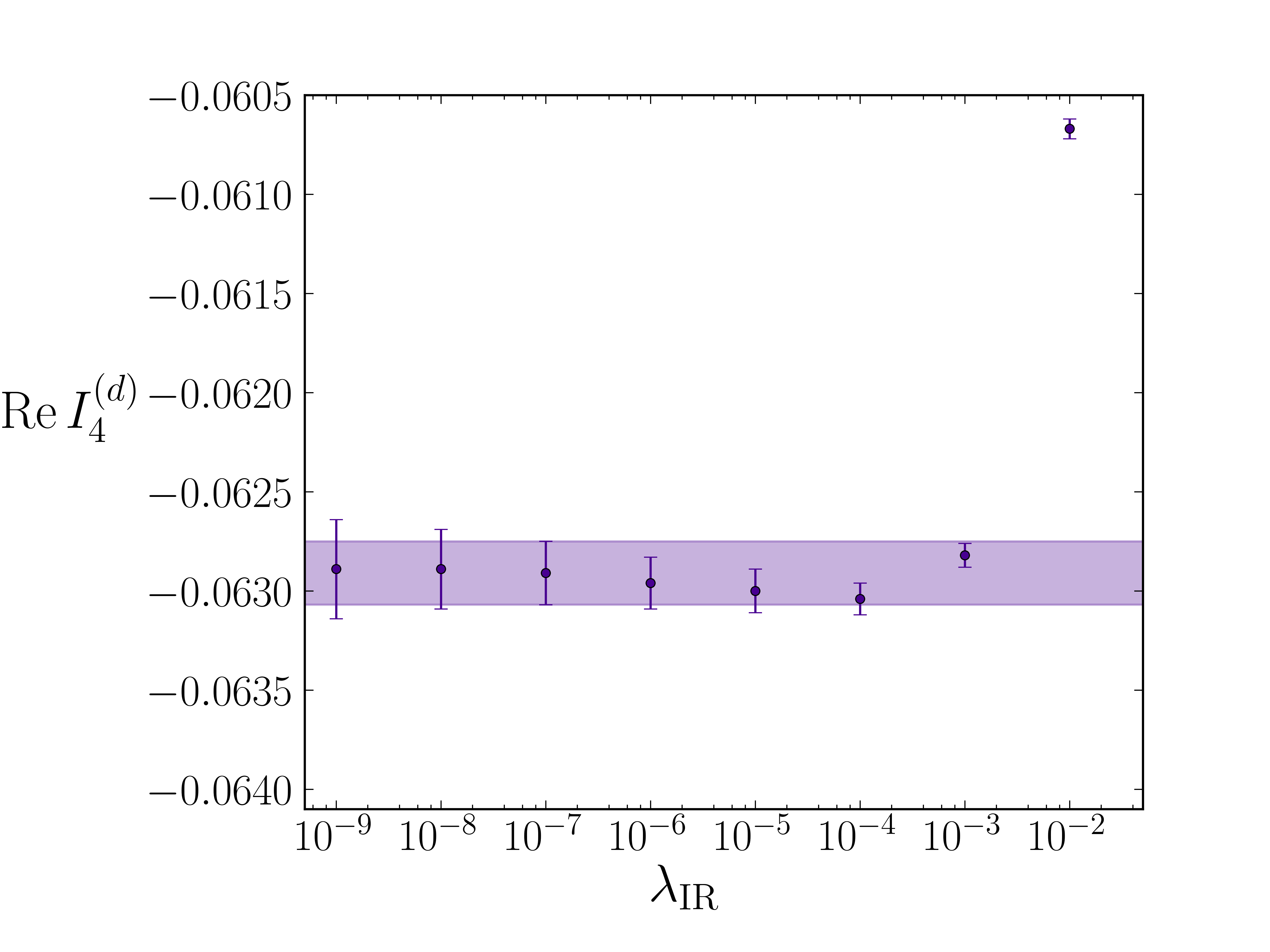}\\
\includegraphics[width=0.4\textwidth,keepaspectratio]{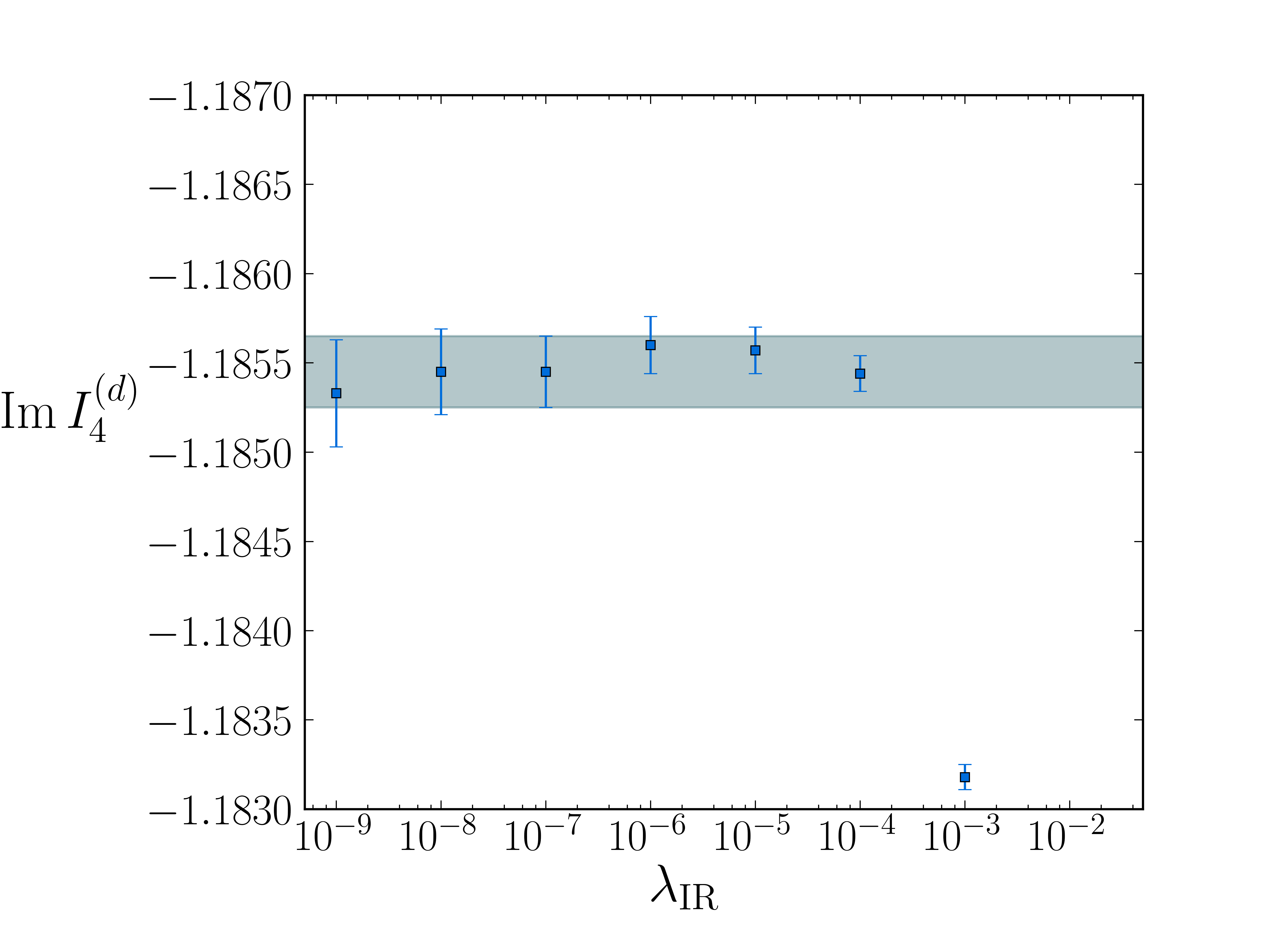}
\end{figure}

Diagrams (d), (e) and (f) suffer from spurious IR instabilities introduced by the numerical evaluation of $(p_z - k_z)$ terms in the denominator. I remove these numerical effects by introducing an IR cutoff $\lambda_{\mathrm{IR}}$, such that momentum points with $|p_z - k_z| \leq \lambda_{\mathrm{IR}}$ are excluded, and demonstrate the numerical convergence of these results, within statistical uncertainties, for the real (upper plot) and imaginary (lower plot) parts of diagram (d) in Fig.~\ref{fig:dcutoff}. For these diagrams, I choose $\lambda_{\mathrm{IR}} = 10^{-7}$, which is illustrated by the shaded band in each plot and is sufficiently small to ensure convergence without unduly increasing statistical uncertainties.

\bibliography{sqpdf_pertth}

\end{document}